\documentclass[english,pre,twocolumn,showpacs,floatfix]{revtex4}
\usepackage[T1]{fontenc}
\usepackage[latin1]{inputenc}
\usepackage{babel}
\usepackage{epsfig}
\usepackage{amsmath}
\usepackage{amssymb}

\begin{document}

\title
{\bf Slowly rocking symmetric, spatially periodic Hamiltonians: The
role of escape and the emergence of giant transient  directed
transport}
\author{ D. Hennig$^{1}$,
  L. Schimansky-Geier$^{1}$, and P. H\"anggi$^{2,3}$}
\medskip
\medskip
\medskip
\affiliation{
  $^{1}$Institut f\"{u}r Physik, Humboldt-Universit\"{a}t
  Berlin,\\Newtonstr. 15, 12489 Berlin, Germany\\
  $^{2}$Institut f\"{u}r Physik, Universit\"{a}t Augsburg,\\
  Universit\"{a}tsstr.~1, 86135 Augsburg, Germany\\
  $^{3}$Department of Physics, National University of Singapore, Singapore
117542, Republic of Singapore}

\begin{abstract}
\noindent The nonintegrable Hamiltonian dynamics of particles placed in a {\it symmetric}, spatially periodic potential and subjected to a periodically varying field is explored. Such systems can exhibit a rich diversity  of unusual transport
features. In particular, depending on the setting of the initial phase of the drive, the possibility of a {\it giant} transient directed transport in a symmetric, space-periodic potential when driven with an {\it adiabatically} varying field arises. Here, we study the escape scenario and corresponding mean escape times of
particles from a trapping region with the subsequent generation of a transient directed flow of an ensemble of particles. It is shown that for adiabatically slow inclination modulations the unidirectional flow proceeds over giant distances.  The direction of escape and, hence, of the flow is entirely governed whether the periodic force, modulating the inclination of the potential, starts out initially positive or negative. In the phase space, this transient directed flow is associated with a long-lasting motion taking place within ballistic channels contained in the non-uniform chaotic layer. We demonstrate that for adiabatic modulations all escaping particles move ballistically into the same direction, leading to a giant directed current.
\end{abstract}

\maketitle

\section{Introduction}\label{intro}
Transport phenomena play a fundamental role in many physical
systems. In this context, dissipative ratchet dynamics has attracted
considerable interest over the past years. Particles placed into
periodic but sawtooth like potentials and driven by external forces
or nonequilibrium noise create a directed flow even if the forces
and noise vanishes in average. Thus unbiased forces induce a
directed motion, a concept which was successfully applied to many
different biological or mesoscopic systems. We refer here to the
various overviews on molecular and Brownian motors
\cite{Ha1996}-\cite{HMN}. This physical concept was subsequently
generalized to potential landscapes  possessing reflection symmetry
which in addition are subjected to asymmetrical driving
\cite{EPL_Luczka}-\cite{Flach} and, as well,  to Hamiltonian
transport in absence of dissipation and enduring agitating
fluctuations \cite{Goychuk}-\cite{Makarov}.

The starting point for our investigation is a system obeying
reflection symmetry, both in space and time. Being interested in the
particle transport features in a nonintegrable situation we study
the following equation of motion
\begin{equation}
  \label{eq:triv1}
  \ddot{q}=F_0(q)+F\,\cos( \Omega t+\Theta_0)=F_0(q)+F(t)\,.
\end{equation}
We consider spatially periodic forces $F_0(q)= F_0(q+1)$ such as
e.g. the one provided by a completely symmetric potential
$U_0(q)=-\cos(2\pi q)/(2\pi)$. The time-varying ac-force in
(\ref{eq:triv1}), $F(t)$, periodically modulates the inclination
of the space-periodic potential $U_0(q)$. This set-up  destroys
the integrability of the system dynamics. In particular,  around
the corresponding separatrix in phase space a chaotic layer
develops. Like in multistable potentials the motion becomes
irregular and trajectories jump erratically from one potential
well to another, being not always the adjacent one. Hence
particles are scattered by the nonlinear forces and, obviously,
the property of a directed current will typically be lost.

Remarkably, as pointed out in  prior literature \cite{Yevtushenko},\cite{GoychukII},
in the system (\ref{eq:triv1}) there results an (unexpected)
asymmetry of the flux of particles, emanating from one potential
well, and flowing to the left and right potential wells which
indicates the existence of directed transport without breaking the
reflection symmetry in space and time in this system. One reason for
the occurrence of phase-dependent directed transport is the lowering
of the symmetry of the flow in phase space by the ac-field where
this asymmetry vanishes only for specific values of the initial
phase $\Theta_0$ \cite{Yevtushenko}.

Also of interest in this context is Ref.~\cite{Soskin}. Therein
the authors report further on this exceptional situation and show
that directed transport is sustained on fairly long time scales
despite the presence of chaos. In particular it has been
demonstrated that for sufficiently small forcing frequencies,
$\Omega \ll 1$, the width of the arising chaotic layer diverges
leading to a strong enhancement of the chaotic transport
\cite{Soskin}.

Here we complement and  extend those prior studies
\cite{Yevtushenko}, \cite{Soskin} to the problem of escape and
successively maintaining {\it exclusive directed motion}. More
precisely, particles from a large ensemble that are initially
contained in a well of the potential have to surmount the
corresponding barrier. We show that adiabatically slow modulations
of the potential lead to that all of the escaping particles not only
leave the potential well in the same direction but subsequently
enter the regime of long-lasting transients for which the transport
proceeds {\it unidirectional} in a ballistic fashion.  The direction
of this arising flow over giant distances depends on the initial
phase $\Theta_0$ about which we do not average.

That trajectories contained initially in the interior of the
separatrix can escape to neighboring wells of the potential is
related with sweeping across the chaotic layer and crossing the
separatrix. Precisely for this situation  the authors in
\cite{Yevtushenko} deomstrated (applying ac-forces with frequencies
$\Omega \sim{\cal{O}}(1)$) that the mean time-averaged velocity of a
particle ensemble decays inversely proportional with time and thus
tending to zero asymptotically. In contrast, we demonstrate here
that for driving with a sufficiently slow ac-field the mean
time-averaged velocity attains a finite value, virtually not
altering on long time scales. As we shall show this is  due to the
fact that there occurs only a single event of separatrix crossing,
namely the one guiding a particle from the inner to the outer
region. Despite that afterwards the particles closely re-approach
the separatrix periodically in time further crossings are avoided.

We further provide reasonings for the occurrence of the flow on the
basis of the underlying phase space structure. The only assumption
we have to make is that particles which are initially trapped in the
interior of the separatrix are included in the arising chaotic layer
around the separatrix in phase space. We will find and discuss the
situation that large ensembles of particles do not only escape from
the separatrix but also move in the same direction. This in turn
yields a giant current.

The paper is organized as follows: In the next section we
introduce the model of the particle motion in a symmetric,
periodic potential, associated with the force. The inclination of
this potential is being time-periodically modulated. The structure of the phase space is elucidated.  In section \ref{sec:layers} 
we investigate the influence of the modulation frequency on the
formation of stochastic layers in phase space. In particular, we
determine the range in which the momentum variable can vary. The analysis in section \ref{sec:growth} 
concerns the
chaos-induced escape of individual trajectories from the interior of the separatrix. In the subsequent section \ref{sec:ratchet}
an explanation for the effect of an enormous enhancement occurring
in the adiabatically driven system is given in terms of the
underlying phase space dynamics. The dynamics of ensembles of
particles contained initially in the interior of the separatrix
and their contributions to a directed flow are addressed in
section \ref{sec:current}. 

We close with a summary of our
obtained results.

\section{The forced nonlinear oscillator model}\label{sec:model}
The equation (\ref{eq:triv1}), as a driven nonlinear Hamiltonian
system in one dimension, is derived from the following Hamiltonian in (dimensionless) form
\begin{equation}
H=\frac{p^{2}}{2}+U(q,t)\equiv\frac{p^{2}}{2}+U_0(q)+U_1(q,t)\,.\label{equation:Hamiltonian}
\end{equation}
Therein $p$ and $q$ denote the canonically conjugate momentum and
position of a particle evolving in the periodic, spatially-symmetric
potential of unit period, i.e.,
\begin{equation}
U_0(q)=U_0(q+1)=-\cos(2\pi q)/(2\pi)\,.
\end{equation}
An external, time-dependent forcing field
\begin{equation}
U_1(q,t)=-F\sin({\Omega \,t+\Theta_0})q
\end{equation}
serves for time-periodic modulations of the inclination of the
potential $U_0$. We underline that the  system is unbiased in the
sense that the force averaged over time and space vanishes, i.e.
\begin{equation}
\int\limits_{0}^{1}dq\int\limits_0^{T=2\pi/\Omega}dt \frac{\partial U(q,t)}{\partial q}=0\,.
\end{equation}

\begin{figure}
\includegraphics[scale=0.3]{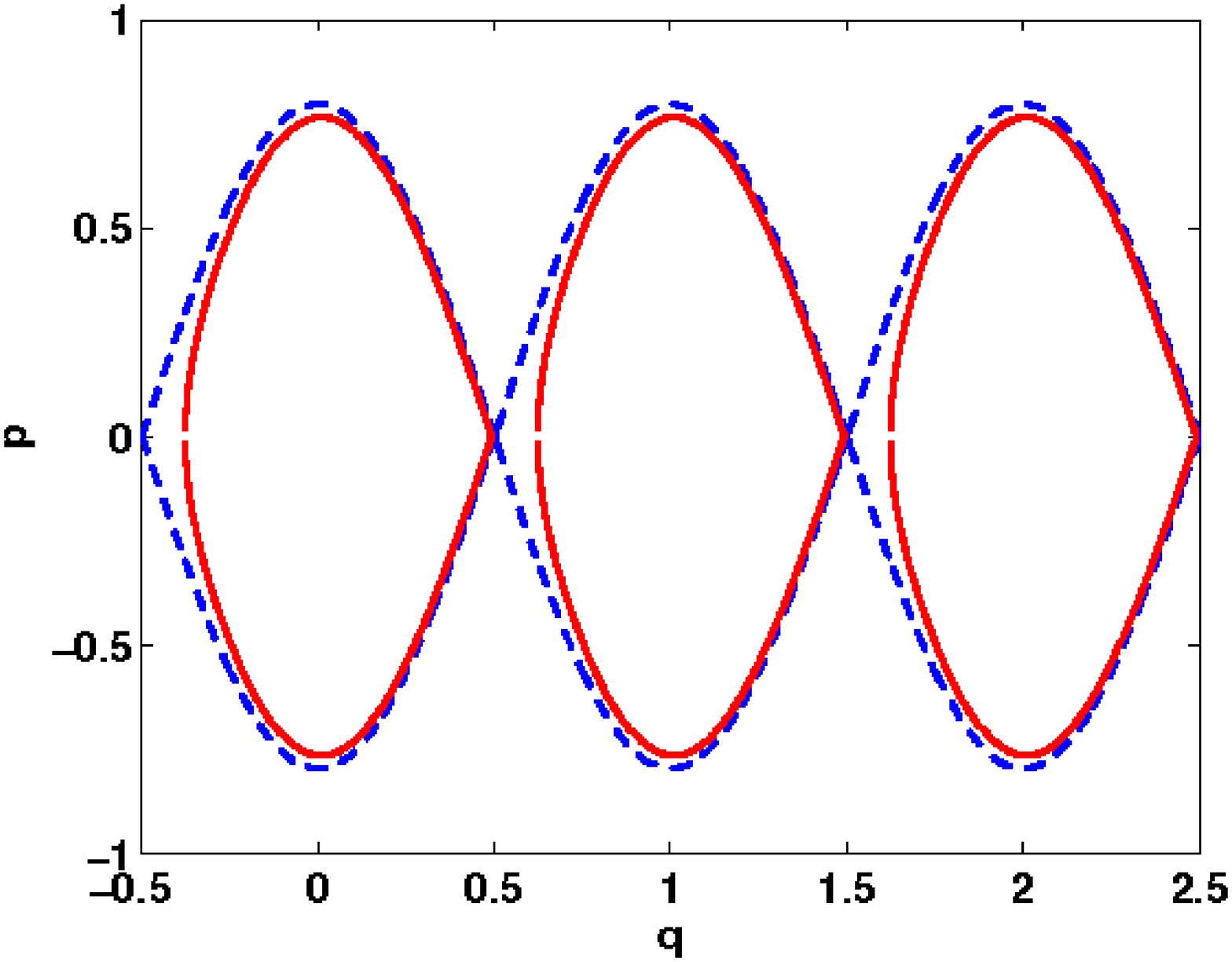}
\includegraphics[scale=0.3]{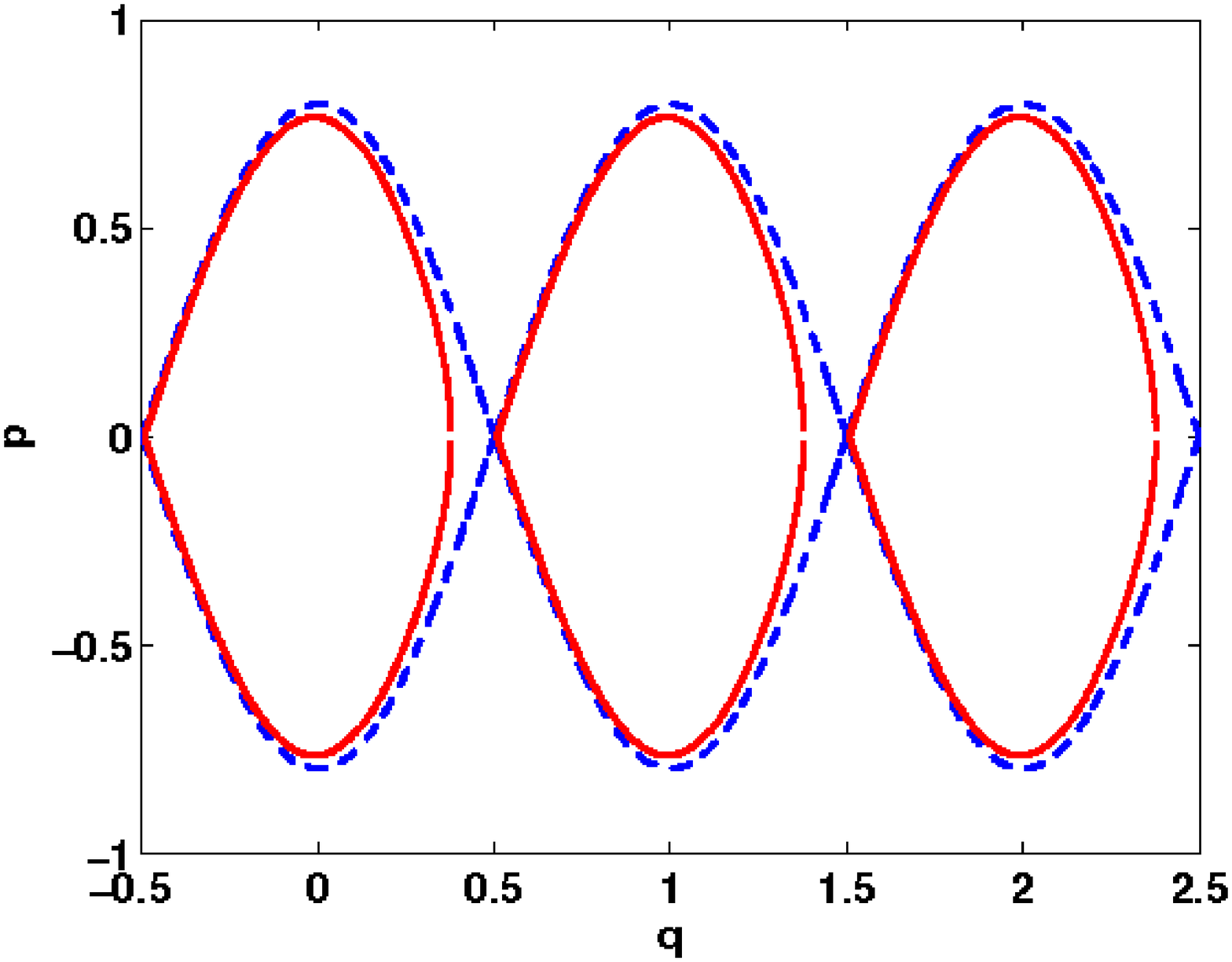}
\caption{(Color online) The separatrix in the phase plane. Top
panel: A chain of homoclinic connections (solid line) for the
potential with negative inclination due to a static force of amplitude
$F=0.05$. The dashed line represents the chain of heteroclinic
connections belonging to a non-inclined potential. Bottom panel: Same
as in the top panel except that now the inclination is reversed, i.e.
$F=-0.05$, implying a positive inclination. The visible small openings in
the solid lines of the separatrix refer to the open channels for
ballistic transport.} \label{fig:potbiased}
\end{figure}

For a static inclination, i.e. for $U_1(q)=-F\,q$ the system dynamics is
integrable and the solutions are contained in the level set
\begin{equation}
H=\frac{p^{2}}{2}-\frac{1}{2\pi}\cos(2\pi q)-F q\equiv E^0\,.
\end{equation}
There exist unstable saddles at ${q}_s^k=0.5+k-\arcsin (F)/(2\pi)$
and stable centers at ${q}_c^k=k-\arcsin (F)/(2\pi)$ with integer
values $k=0,\,\pm1, \pm 2\,\,...$. In the non-inclined case,
$F=0$, neighboring hyperbolic points are linked via heteroclinic
connections and with inclination-modulation $F \ne 0$ each
hyperbolic point is linked with itself via a homoclinic connection
as illustrated in Fig.~\ref{fig:potbiased}. According to the
location of the saddles to the right (left) of the centers we
denote the chain of homoclinic connections in the left (right)
panel as right-oriented (left-oriented). For later reference, we
point to the open channel arising between two neighboring
separatrix loops for inclined potentials.
The separatrix energy attributed to the saddle at ${q}_{s}^k$ is given
by
\begin{equation}
E_{separatrix}^0=\frac{1}{2\pi}\left[\sqrt{1-F^2}-F[(0.5+k)2\pi-\arcsin(F)]\right].
\end{equation}
The maximal extension in momentum of the separatrix is
\begin{equation}
p_{max}^{\pm}=\pm\left(\frac{2}{\pi}\sqrt{1-F^2}\right)^{1/2}\,,
\end{equation}
and the separatrix crosses the $q-$axis (turning point of the
corresponding homoclinic orbit) at a value $q_0$ that is
determined by the solution of the transcendental equation
\begin{equation}
E_{separatrix}^0+\frac{1}{2\pi}\cos(2\pi q_0)+F q_0=0\,.
\end{equation}
Note that the separatrix loops for $F=0$ completely comprises
those for $F\ne 0$. For time-periodic modulations of the
inclination which are imposed by the time-dependent potential
$U_1(q,t)$, the phase space structure becomes governed by
breathing of the separatrix loops where the left-oriented and
right-oriented phase space structures displayed in
Fig.~\ref{fig:potbiased} represent the "turning points" of the
breathing. In-between these turnings the area enclosed by a
separatrix loop is periodically changing and the minimum and
maximum is obtained when $\Omega t+\Theta_0=\pi/4,3\pi/4$ and
$\Omega t+\Theta_0=0,\pi,2\pi$ respectively. Moreover, with the
application of a time-dependent field $U_1(q,t)$ a breaking of the
integrability of the dynamics is expected. In particular, around
the separatrix of the unperturbed system a chaotic layer forms.
These oscillations of the separatrix between the left-oriented and
right-oriented structures have to be distinguished from the
pulsations of the width of the pendulum separatrix considered in
\cite{Menyuk,Cary,Elskens}. In Ref.~\cite{Coppola} the chaos
dynamics in a system with periodically disappearing separatrix was
considered.

\section{Chaotic layers and Poincar\'{e}-plots}
\label{sec:layers}
\begin{figure}
\includegraphics[scale=0.3]{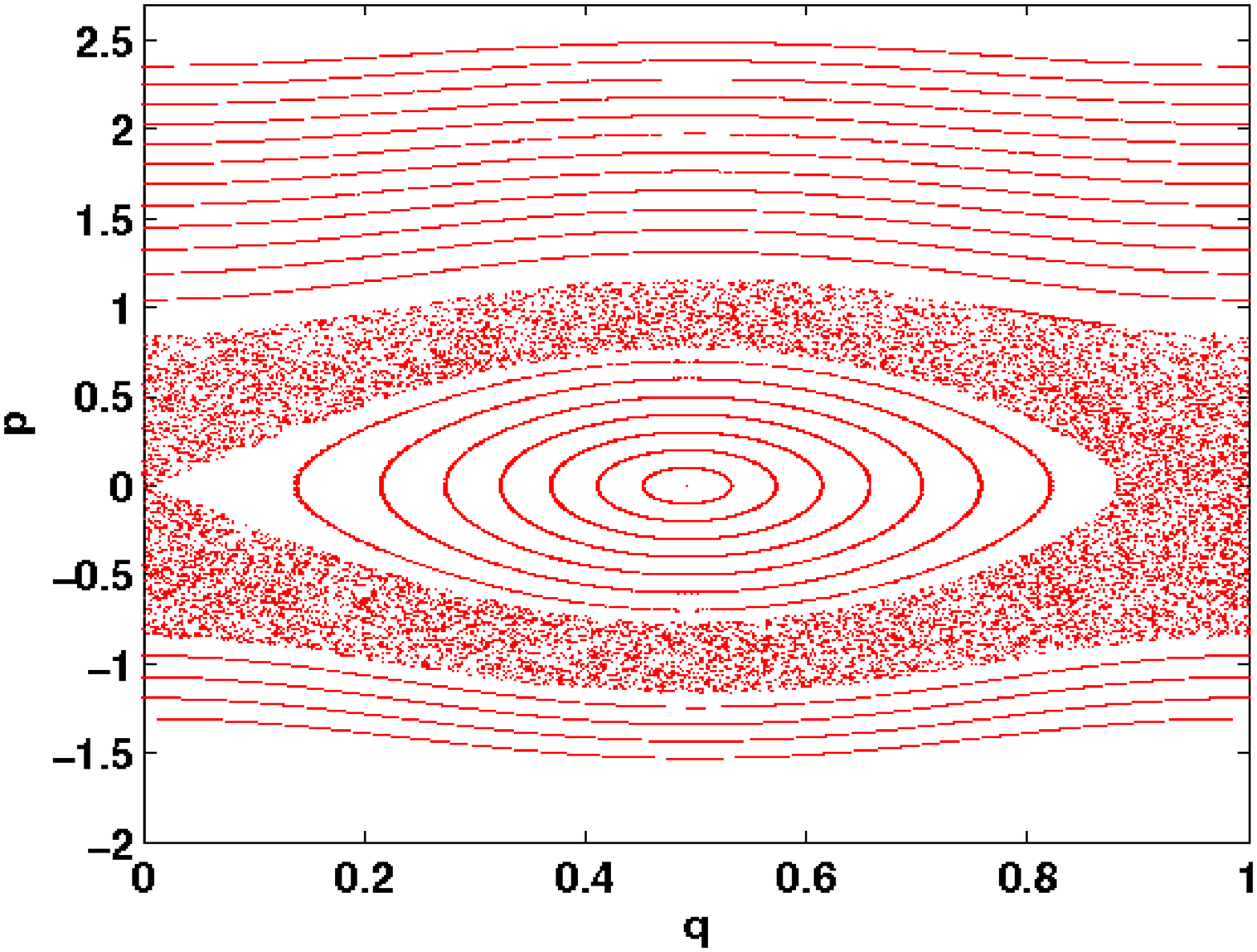}
\includegraphics[scale=0.3]{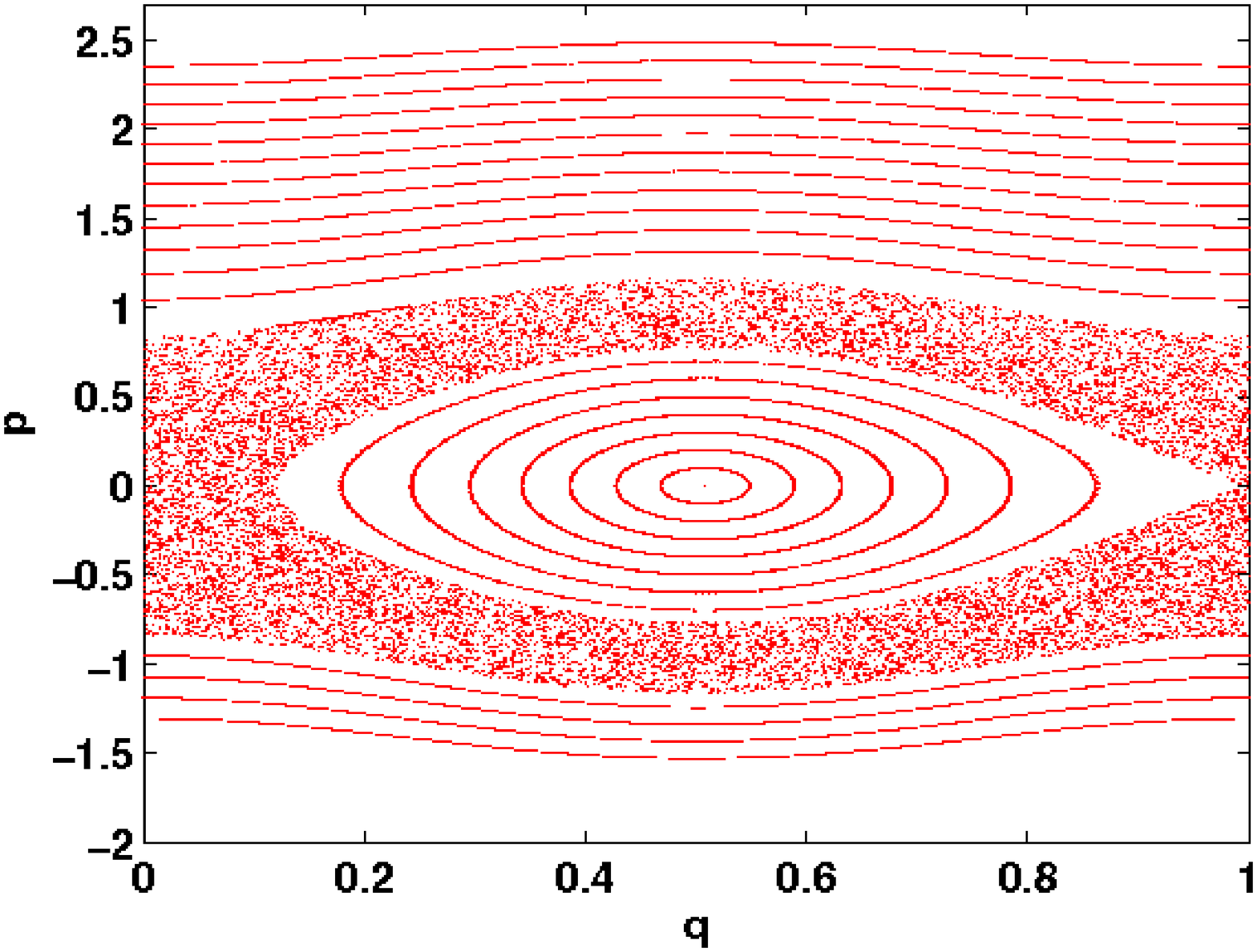}
\caption{(Color online) Stroboscopic Poincar\'{e}-plots for an
angular frequency $\Omega=0.1$, force strength $F=0.05$ and
initial phase $\Theta_0=0$. The top and the bottom panel are taken
at $t_k=\left(\frac{1}{4}+k\right)T$ (maximal negative
inclination) and $t_k=\left(\frac{3}{4}+k\right)T$ (maximal
positive inclination), respectively. The coordinate $q$ is
represented $\mod 1$.} \label{fig:poincare1}
\end{figure}

\begin{figure}
\includegraphics[scale=0.3]{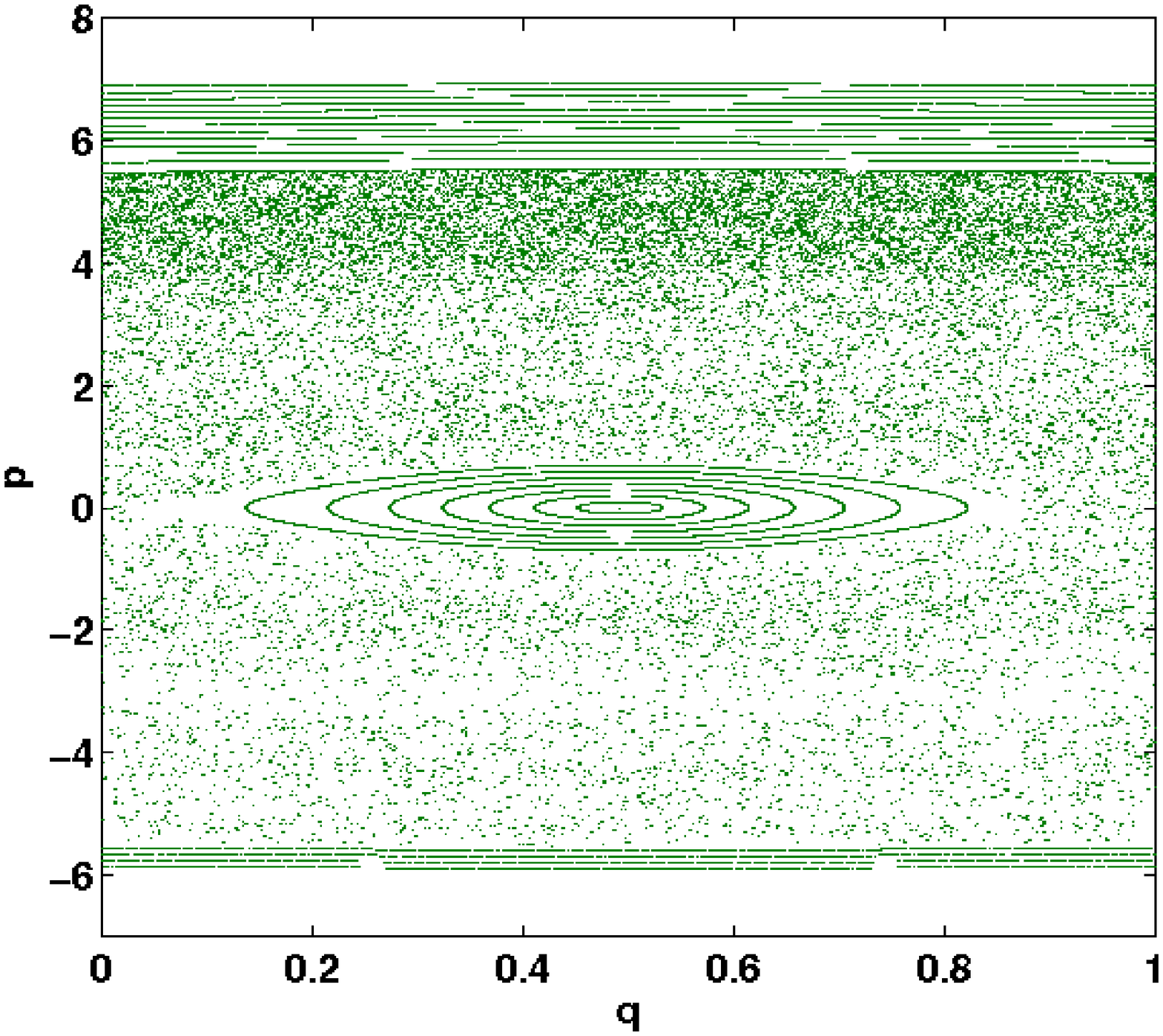}
\includegraphics[scale=0.3]{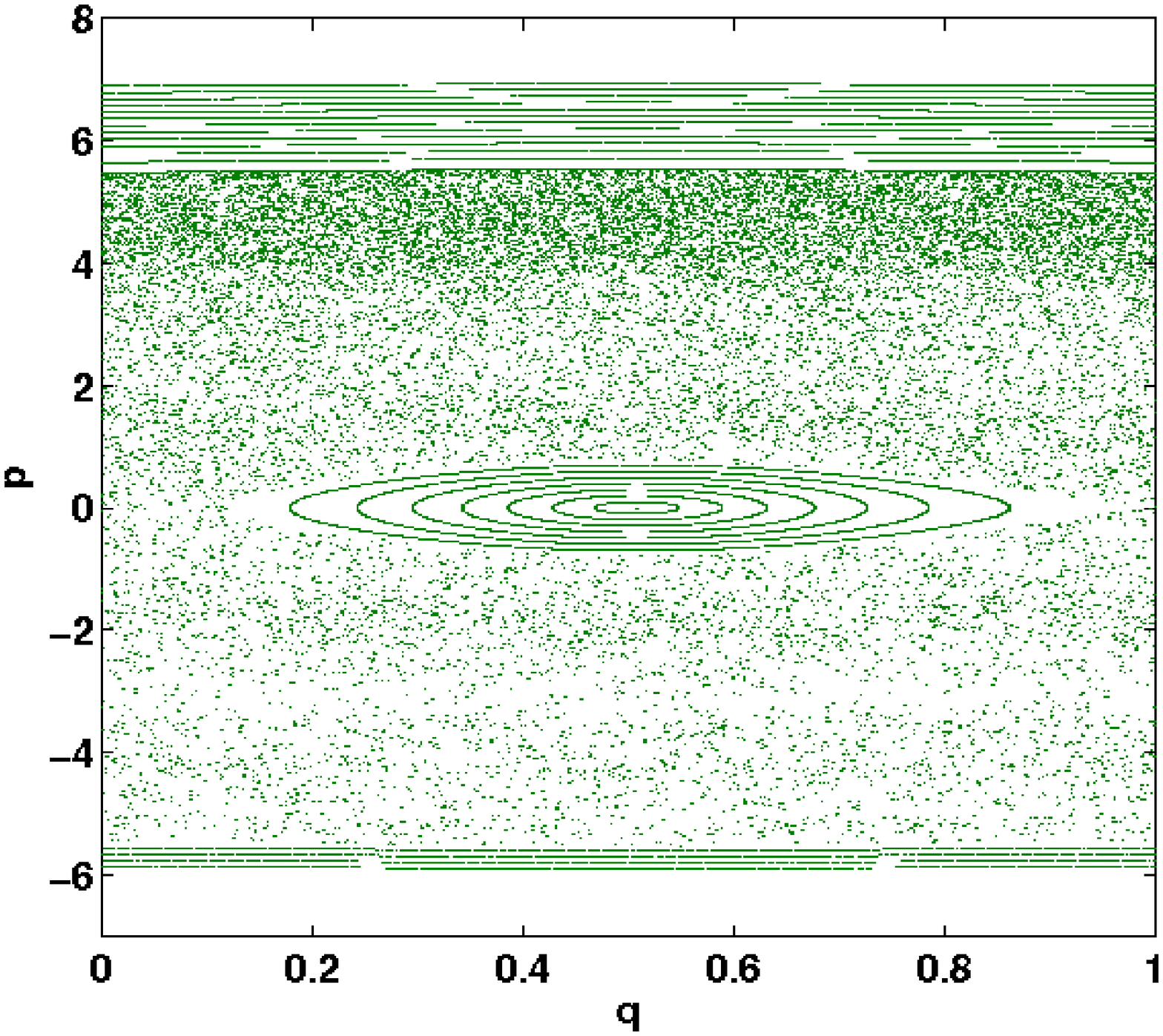}
\caption{(Color online) 
Stroboscopic Poincar\'{e}-plots for an
angular frequency $\Omega=0.01$, force strength $F=0.05$ and
initial phase $\Theta_0=0$. The top and the bottom panel are taken
at $t_k=\left(\frac{1}{4}+k\right)T$ (maximal negative
inclination) and $t_k=\left(\frac{3}{4}+k\right)T$ (maximal
positive inclination), respectively. The coordinate $q$ is
represented $\mod 1$.}
\label{fig:poincare3}
\end{figure}

To illustrate the influence of the angular driving frequency on the
behavior of the system we depict in Figs.~\ref{fig:poincare1} and
\ref{fig:poincare3}
the stroboscopic Poincar\'{e}-plots at successive periods of the
driving term $T= 2\pi/\Omega$ in the $p-q-$plane for force
strength $F=0.05$, phase $\Theta_0=0$ and two different driving
frequencies $\Omega=0.1$ and $\Omega=0.01$. The simulation time is
chosen such that $\Omega\, t=10^4$ holds and therefore, the number
of periodic changes $10^4/\pi$ of the inclination of the potential
in the simulation time interval is the same for any frequency.

The top panel in Figs.~\ref{fig:poincare1} and \ref{fig:poincare3}
corresponds to stroboscopic plots at $t_k=(1/4+k)T$ with
$k=0,1,2,...$ where the potential assumes a maximal negative
inclination. The bottom panel belongs to stroboscopic plots at
$t_k=(3/4+k)T$, i.e. the potential has a maximal positive
inclination. For fairly fast varying modulations with $\Omega=1$ a
chaotic layer of small width develops around the separatrix,
whereas the motion remains trapped and regular within the large
island of stability with its center at the origin of the phase
plane. The deformed horizontal lines above and below the large
island of stability correspond to KAM tori which act as barriers
for transport impeding larger upwards and downwards excursions of
the momentum variable. Note that the extension of the chaotic
layer in momentum direction remains equal regardless of the sign
of the inclination of the potential.

Nevertheless, for slow modulation with $\Omega=0.01$ this picture
changes drastically. Many KAM tori become destroyed and the only
surviving ones lie in the region of large 
$|p| \gtrsim 5.8$. There
remains still an island of stability corresponding to bounded and
regular motion inside the potential well. On the other hand, the chaotic
layer has grown considerably in momentum direction compared to the previous case of $\Omega=1$.
Most strikingly, independent of the
sign of the inclination of the potential
the density of the points in the stroboscopic plots is much higher in the region
of positive momentum than in the region of negative ones. Thus, the momentum
variable is allowed to raise to fairly large positive values. Due
to symmetry for an initial phase value $\Theta_0 =\pi$ of the
external field, i.e. when during the first half-period $0<t<T/2$
the inclination becomes positive, an equivalent behavior is
observed except that the cloud of points
now penetrates more into
the range of negative $p$.

In the following we focus our interest on the generation of a
directed flow for a large ensemble of particles
where {\it all of them} perform {\it exclusively ballistic} motion
in the same direction. Note that this represents a  far stronger
condition than merely observing directed, diffusive transport of particles as
illustrated in \cite{Yevtushenko}.
Our set up is the following: If the potential assumes, let us say,
a negative inclination due to a static force $-\partial
U_1/\partial q=F$ we suppose that the dynamics of the particles is
bounded in one well and the corresponding potential barrier is
insurmountable, i.e. for all particles it holds that
$E_{particle}^0=H_0=p^2/2+U_0(q)<E_{separatrix}^0$. Accordingly,
the initial conditions for our simulations are distributed in the
interior of the corresponding separatrix loop (see also below in
Fig.~\ref{fig:escapeset}).  Obviously, with static inclination of
the potential the integrable dynamics is characterized by
oscillations around the stable elliptic center in the separatrix
loop and hence, the particles remain trapped in the potential
well. However, this scenario drastically changes when the
time-dependent perturbations destroy the integrability. In
particular homoclinic chaos is present in the driven dynamics. In
fact, those trajectories seized by the arising chaotic layer may
manage to escape from the interior region of the broken
separatrix. We emphasize that this trapping-detrapping transition
can only be triggered by the chaotic dynamics within the chaotic
layer.

When applying the time-periodic modulation of the inclination with
initial phase set at $\Theta_0=0$ our starting point is a
non-inclined potential. Motion is then supposed to proceed towards
the right. Since in the initial stage $0<t<T/2$, the inclination
passes from zero value to its maximal negative value it
contributes to a continually growing positive force which promotes
the desired motion towards the right.  The question then is: What
determines an efficient escape of trapped particles starting out
from distributed initial conditions? Is the direction of the
motion of the escaped particles determined completely by the
choice of the phase of the modulation term? If so, for which
parameter values do the escaped particles keep moving (on average)
in the preferred right direction on longer time scales ?

\section{Separatrix crossing and energy variations}\label{sec:growth}
In this section we discuss the trajectories of particles escaping
from the interior of the separatrix loop.  Furthermore, the
possibility of energy growth after separatrix crossing for slow
modulations of the inclination is studied.

\subsection*{Separatrix crossing and escape}
\begin{figure}
\includegraphics[scale=0.4]{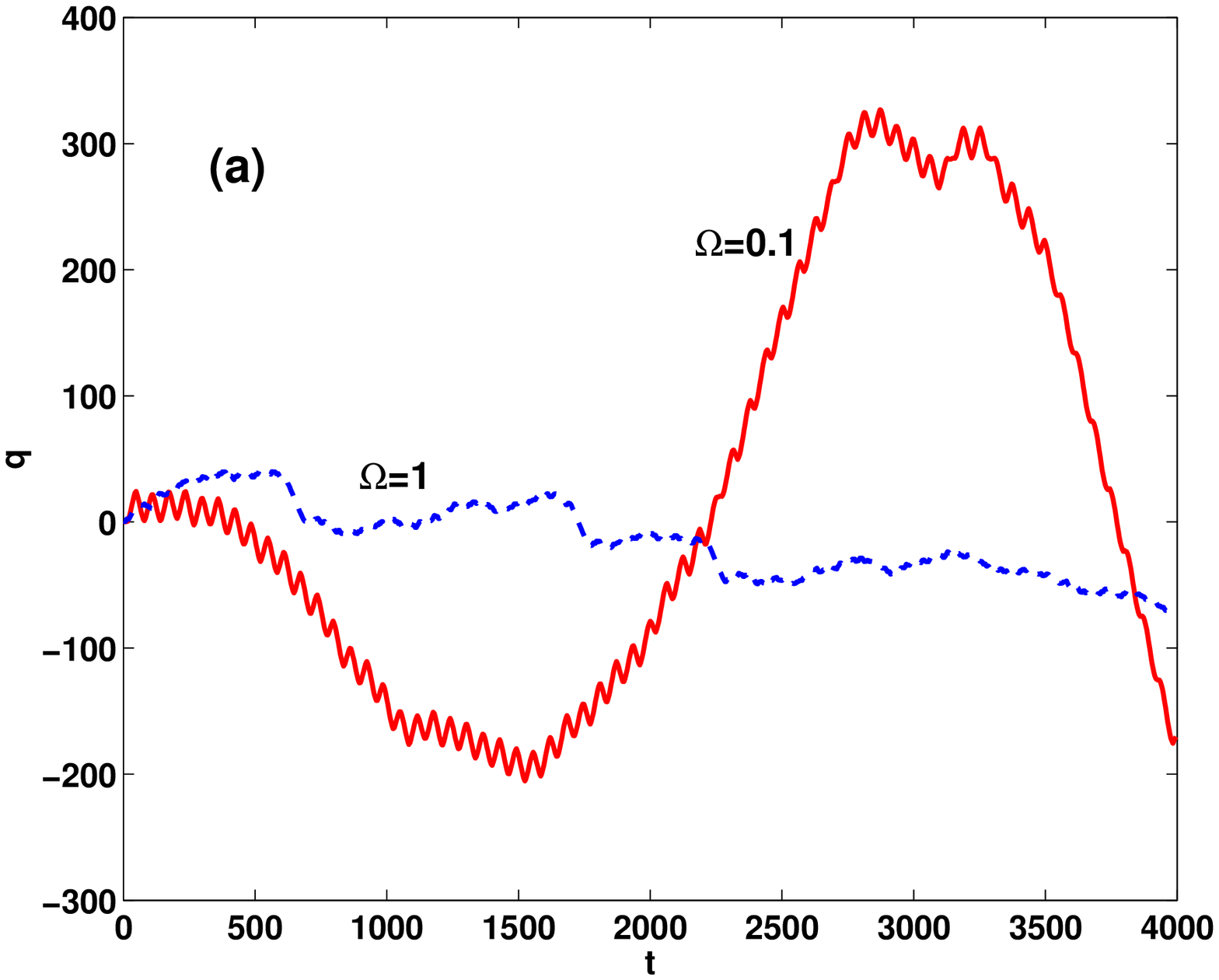}
\includegraphics[scale=0.4]{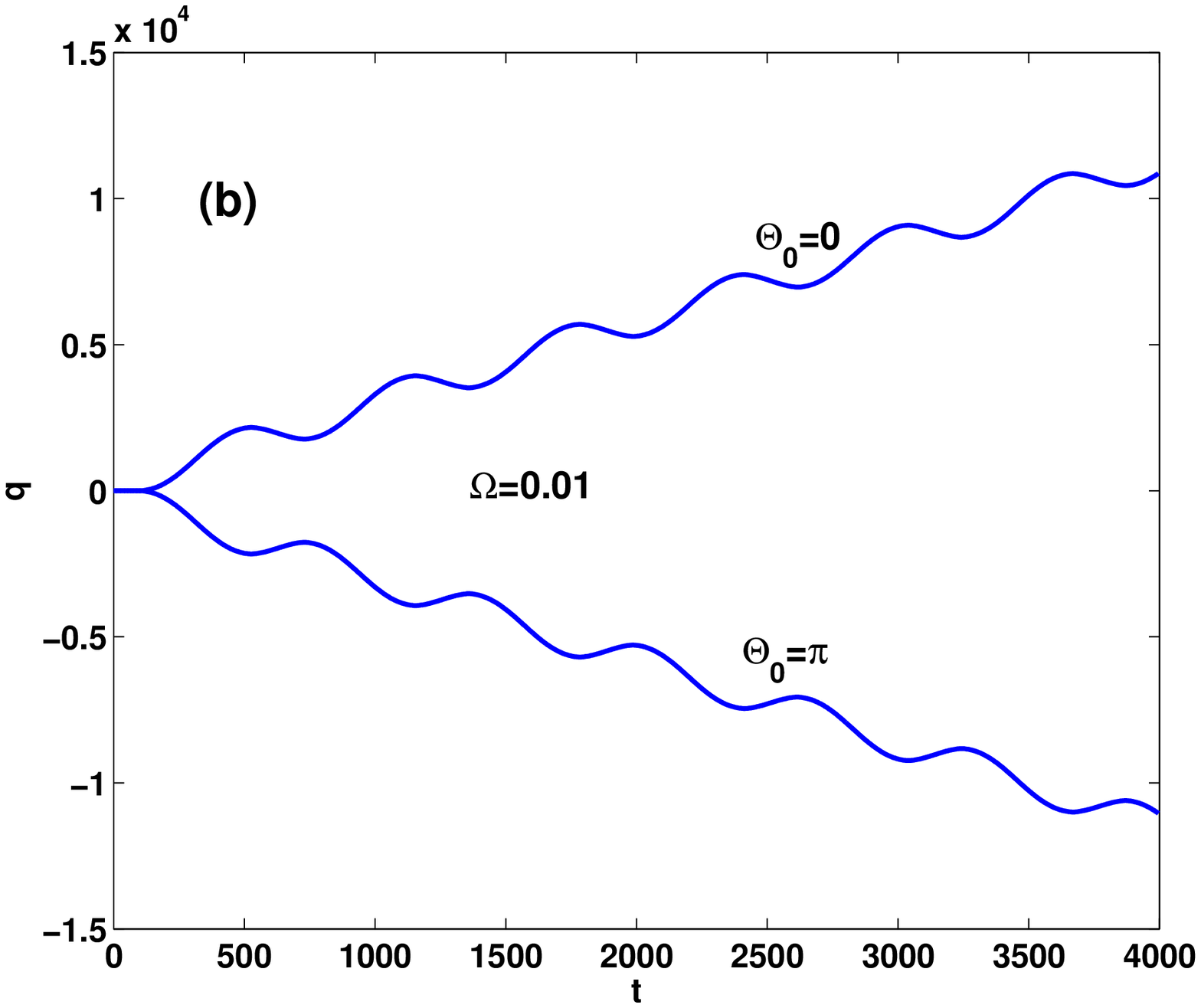}
\includegraphics[scale=0.4]{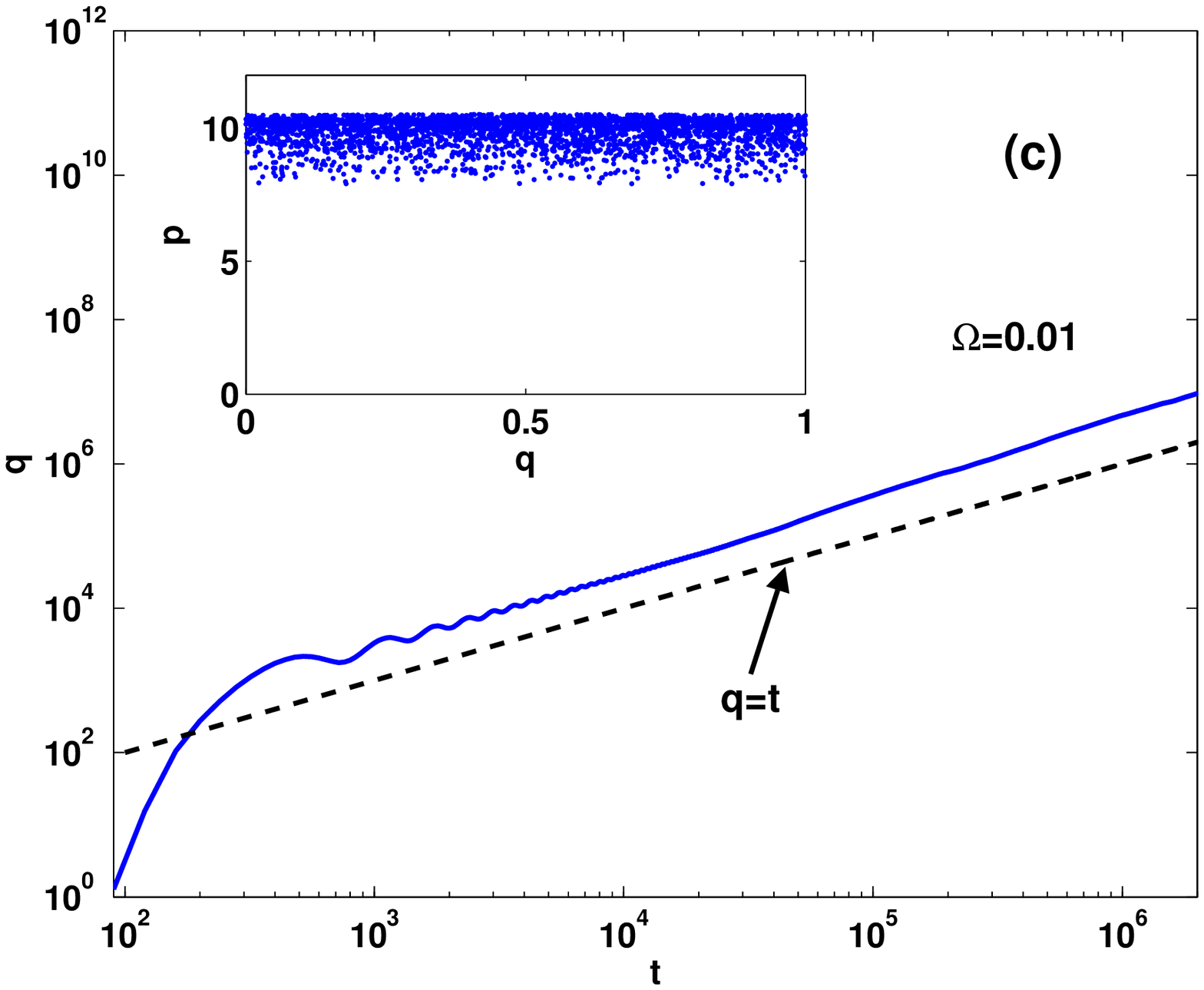}
\caption{(Color online) Time evolution of the coordinate $q(t)$
depicted with panels (a-c) for
  three different angular driving frequencies $\Omega$ (indicated in
  the corresponding panels (a)-(c)) but equally chosen initial conditions $p(0)=0$ and
  $q(0)=0.423$. The remaining parameter values are $F=0.05$ and
  $\Theta_0=0$.  In the central panel (b) we also depict the trajectory for
  initial coordinate $q(0)=-0.423$ and phase $\Theta_0=\pi$ for which
  the motion proceeds in the negative direction. In the bottom panel (c) the
  long-lasting ballistic motion of the particle is illustrated. Note
  the large values for the simulation times  $ > 10^6 $ in panel (c). The inset depicts a stroboscopic Poincar\'{e}
  plot (at $t_k=Tk$) of the trajectory, revealing the motion in a
  ballistic channel.}
\label{fig:q1.fig}
\end{figure}

In Figs.~\ref{fig:q1.fig} we depict the role of the angular frequency
on the time evolution of trajectories for an initial condition
contained in the region in the interior of the unperturbed separatrix
in which the chaotic layer arises. For moderate and intermediate
angular driving frequencies $\Omega=1$ and $\Omega=0.1$, respectively,
the coordinate $q(t)$ behaves chaotically, exhibiting sudden and
unpredictable changes of direction.

In clear contrast, for slow driving at $\Omega=0.01$ the coordinate
dynamics not only seemingly behaves more regular but also grows on
average upon evolving time. Due to symmetry the direction of the
motion is reversed by changing the initial position $q(0)$ to $-q(0)$
and taking for the phase $\Theta_0=\pi$.

Furthermore, there are alternating long and short intervals during
which the particle moves straightforwardly towards the right and left,
respectively. Correspondingly, the momentum $p(t)$ evolves in phases
with positive value (motion towards the right) that are longer lasting
than the phases when the momentum is negative (leftwards motion) (see
also further below in Fig.~\ref{fig:mechanism.fig}). As
Fig.~\ref{fig:q1.fig} also reveals, the directed motion is maintained
on a very long time scale where the coordinate assumes huge values.
Further details concerning the time scale of unidirectional motion are contained in  Sec.~\ref{sec:current}.
In the bottom panel of Fig.~\ref{fig:q1.fig} one recognizes that for
times $t \gtrsim 10^3$ the trajectory is trapped in a {\it ballistic
  channel} where the particle behaves effectively like a free particle
propagating ballistically. Notice that the simulation time interval
$T_s=2\times10^6$ is equivalent to almost $8\times10^5$ and $3184$
times the period duration for harmonic oscillations near the bottom of
a potential well and the external modulation, respectively.

Such motion in a ballistic channel occurs for the complex dynamics
of systems with a mixed phase space
\cite{Zaslavsky}-\cite{Denisov3}. In more detail the (broad)
chaotic layer is not uniform and contains cantori which can
severely restrict the transport in phase space and thus
effectively partition the chaotic layer bounded from below and
above by non-contractible KAM tori \cite{MacKay}-\cite{Meiss}.
Islands of regular motion that are situated at the upper and lower
boundary of the layer in the vicinity of the confining KAM tori
possess non-zero winding numbers and thus facilitate transport.
The motion around these islands is characterized by the stickiness
to them \cite{MacKay}-\cite{Meiss}, that can lead to trapping of
the trajectory for a long time resulting in ballistic motion
\cite{Denisov2,Denisov3}. For some islands the sticking times to
the boundary of the islands can be anomalously long
\cite{Shlesinger,Klafter}.

\subsection*{Energy variations}
It is illustrative to consider the difference,
\begin{equation}
\Delta E =E_{particle}-E_{separatrix}\label{equation:energetic}
\end{equation}
between the particle energy
\begin{equation}
E_{particle}=\frac{p^2}{2}+U_0(q)-F\sin(\Theta)q\,,\label{equation:eparticle}
\end{equation}
and the energy of the "frozen" separatrix at time $t$
\begin{equation}
E_{separatrix}=U_0({q}_s^k)-F\sin(\Theta) {q}_s^k\,,
\label{equation:endifference}
\end{equation}
where $\Theta=\Omega t+\Theta_0$ and ${q}_s^k=0.5+k+\arcsin
(F\sin(\Theta))/(2\pi)$
denote the instantaneous position of the
corresponding hyperbolic point when the trajectory traverses the
actual range
\begin{eqnarray}
{q}_s^{k-1}\le q \le {q}_s^{k}\qquad &{\rm if}&\qquad  p>0\nonumber\\
{q}_s^{k}\le q \le {q}_s^{k+1}\qquad &{\rm if}& \qquad p<0\,.
\end{eqnarray}
(Note that as $q$ is a dynamical variable so is $k$.)
\begin{figure}
\includegraphics[scale=0.45]{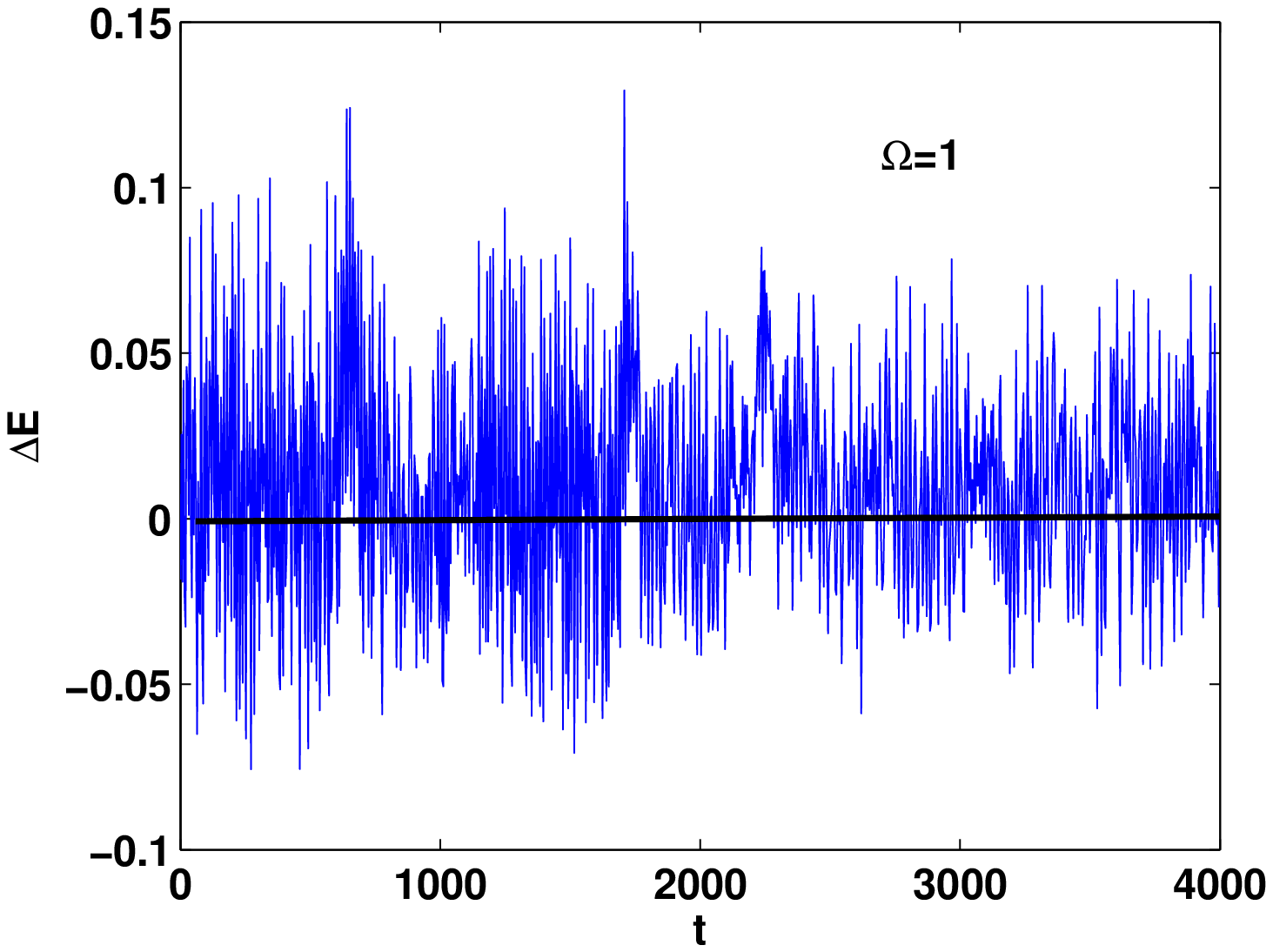}
\includegraphics[scale=0.45]{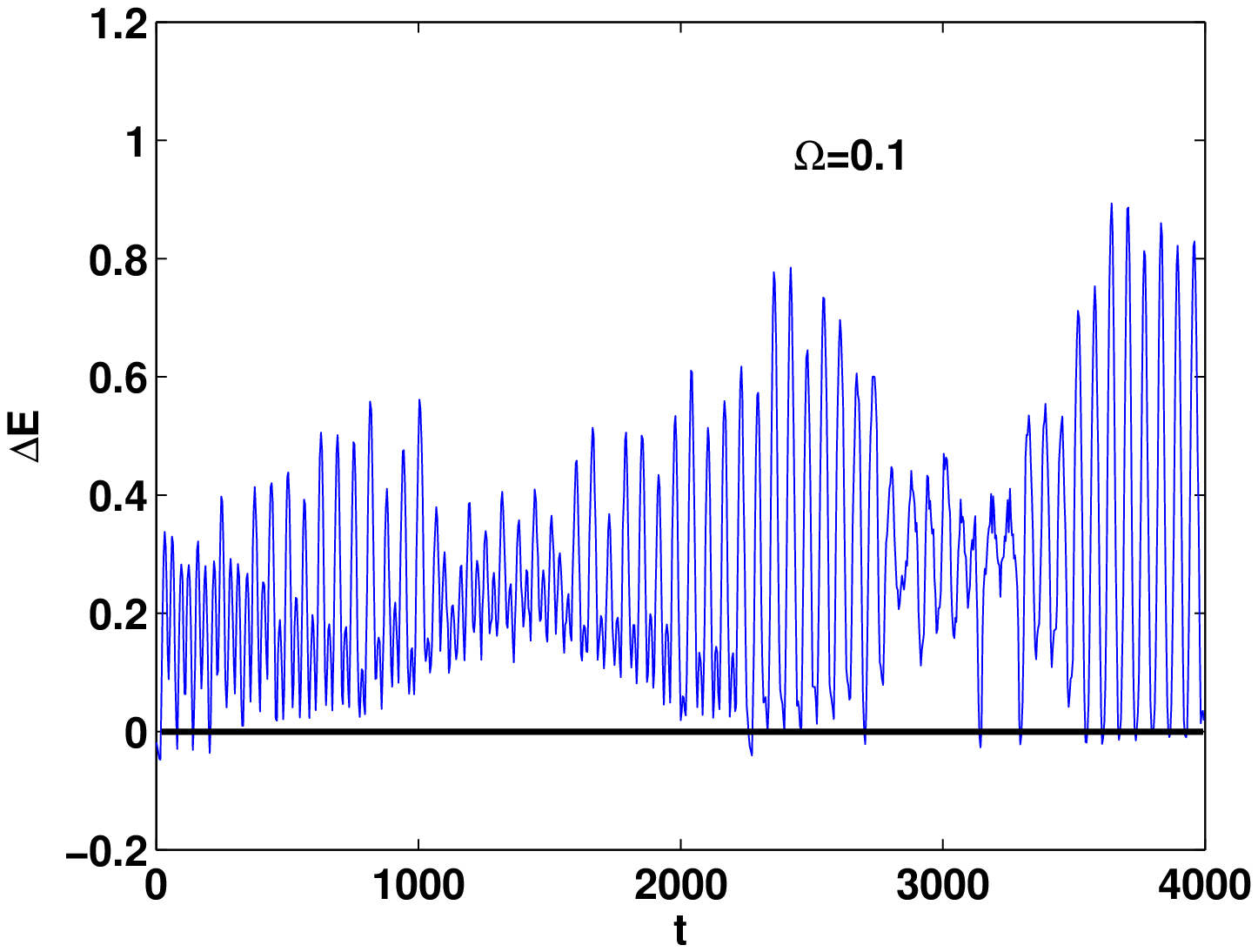}
\includegraphics[scale=0.45]{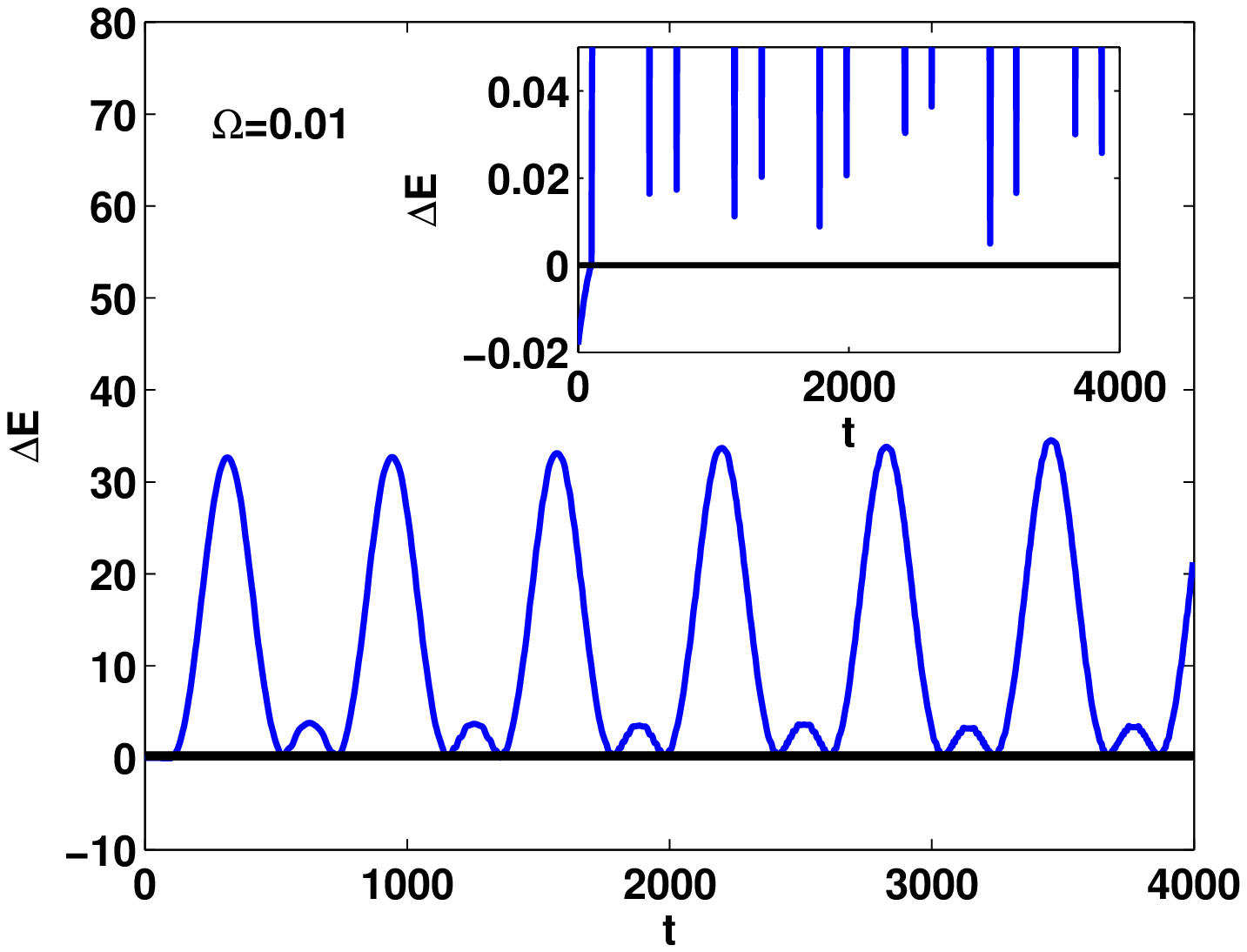}
\caption{(Color online) Temporal behavior of the energy difference
$\Delta E$ as defined in (\ref{equation:energetic}) for three
different angular driving frequencies $\Omega$ 
(as indicated in
  the top, central and bottom panel) but equally chosen initial conditions $p(0)=0$ and
  $q(0)=0.423$. The remaining parameter values are $F=0.05$ and $\Theta_0=0$. 
In the inset in the bottom panel one recognizes that only a single separatrix
crossing takes place.} \label{fig:sep1.fig}
\end{figure}

As Fig.~\ref{fig:sep1.fig} reveals, for a fast modulation
$\Omega=1$ the energetic difference $\Delta E$ changes frequently
the sign, corresponding to the trajectories' repeated leaving and
re-entering of the interior of the instantaneous separatrix.
Nevertheless, in-between separatrix crossings, the particle can be
trapped in a potential well for some time. Since $\Delta E$  stays
close to zero the trajectory remains close to the separatrix for
most of the time. Decreasing the angular driving frequency to
$\Omega=0.1$ has the effect that the number of separatrix
crossings diminishes. Occasionally there appear interludes during
which the trajectory escapes from the separatrix region giving
rise to considerable coordinate changes  $q$ (cf.
Fig.~\ref{fig:q1.fig}). However, these changes are not
coordinated, so that in essence no directed motion results.
\begin{figure}
\includegraphics[scale=0.5]{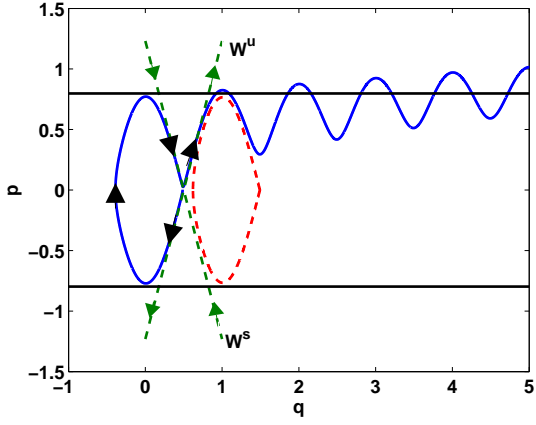}
\caption{(Color online)
Illustration of the escape from the region of the hyperbolic point
in the phase plane. The particle gathers further momentum after the
final libration in the interior of the separatrix has taken place
(oscillatory solid line). The straight dashed lines run in the
directions of the eigenvectors belonging to the eigenvalues of the
system linearized about the hyperbolic point. Sufficiently close to
the hyperbolic point they represent the stable and unstable
manifold $W^s$ and $W^u$. The two horizontal solid lines at
$p_{\pm}=\pm \sqrt{2/\pi}$ confine
the range of the momenta variations of the unperturbed separatrix.
The dashed closed line to the right of the hyperbolic point
represents the instantaneous separatrix loop frozen at the moment
when $p$ attains the value $p_+$. The parameter values are given
by $F=0.05$, $\Omega=0.01$ and $\Theta_0=0$.} \label{fig:pqesc.fig}
\end{figure}

This behavior drastically differs from that occurring for a slow
modulation $\Omega =0.01$. Once the trajectory has crossed the
separatrix it gathers momentum as long as the inclination remains
non-positive. This goes along with an increase in the energetic
difference $\Delta E$. Afterwards, during the depreciation period,
i.e. when the inclination of the potential is positive, the
trajectory moves back towards the instantaneous separatrix, but
never actually re-enters its interior region. Notably, we followed the
evolution on time scales as long as $10^7$ where this behavior
still holds true. In other words reversions of the direction of
motion proceed in an open channel between two adjacent separatrix
loops of the inclined potential as long as the {\it adiabaticity
condition}
\begin{equation}
T_{trajectory} \ll
T=\frac{2\pi}{\Omega}\,.\label{equation:adiabatic}
\end{equation}
is satisfied, where for near-separatrix motion the period duration
$T_{trajectory}$ is asymptotically determined by
\begin{equation}
T_{trajectory}=\frac{4}{\sqrt{2\pi}}\ln\left[\frac{4}{\sqrt{\pi|E^0_{particle}-E^0_{separatrix}|}}\right]\,.
\end{equation}
For the evolution
depicted in the bottom panel in Fig.~\ref{fig:q1.fig} and Fig.~\ref{fig:sep1.fig}
this condition (\ref{equation:adiabatic})  is obeyed.

The influence of the period of the driving
force on the escape of the trajectory through the
saddle point region can be elucidated as follows: Near the saddle point ${q}_s$ the dynamics is
given by
\begin{equation}
\frac{d^2q}{dt^2}\,+\,U_0^{\prime\prime} ({q}_s)q \,=\,F\sin(\Omega
t +\Theta_0)\,,\label{equation:linear}
\end{equation}
where
\begin{equation}
U_0^{\prime\prime}({q}_s)=\frac{d^2U_0}{dq^2}_{\rvert_{q={q}_s}}
=-2\pi\equiv -a\,.
\end{equation}
The solution of equation (\ref{equation:linear}) with initial
condition $p(0)$ and $q(0)$ is given by
\begin{eqnarray}
p(t)&=&\left[\,p(0)+F\cos(\Theta_0)\frac{\Omega}{a+\Omega^2}\,\right]\cosh(\sqrt{a}t)\nonumber\\
&+&\sqrt{a}\left[q(0)+F\frac{\sin(\Theta_0)}{a+\Omega^2}\,\right]\sinh(\sqrt{a}t)\nonumber\\
&-&F\frac{\Omega}{a+\Omega^2}\cos(\Omega
t+\Theta_0)\,\label{equation:plinear}\\
q(t)&=&\frac{1}{\sqrt{a}}\left[\,p(0)+F\cos(\Theta_0)\frac{\Omega}{a+\Omega^2}\,\right]\sinh(\sqrt{a}t)\nonumber\\
&+&\left[q(0)+F\frac{\sin{\Theta_0}}{a+\Omega^2}\,\right]\cosh(\sqrt{a}t)\nonumber\\
&-&F\frac{1}{a+\Omega^2}\sin(\Omega
t+\Theta_0)\,.\label{equation:qlinear}
\end{eqnarray}
For directed motion to occur it is important  that the trajectory,
after having crossed the separatrix,  gathers  enough momentum
that a sufficient distance to the separatrix attributed to the
saddle point(s) of the unstable equilibrium of the next adjacent
potential well(s) is reached before the inclination of the
potential is reversed. Such behavior is illustrated with
Fig.~\ref{fig:pqesc.fig} for slow driving with $\Omega=0.01$.

From the behavior of the solution
(\ref{equation:plinear}),(\ref{equation:qlinear}) with initial
conditions situated close to the hyperbolic point we find that for
relatively fast driving with $\Omega=1$ the distance of the
trajectory immediately grows  because of the still rapidly
decreasing inclination of the potential being connected with
increasing momentum and coordinate of the particle. Rather soon
the sign of potential inclination, and thus, the direction of the
particle motion are reversed
while the trajectory has not departed from the region close to the hyperbolic point.
Correspondingly the trajectory,
whilst being still in the neighborhood of the hyperbolic point,
approaches the nearby part of the adjacent
separatrix (represented by the dashed line in Fig.~\ref{fig:pqesc.fig})
and eventually crosses it soon after the escape.
Contrarily, for slower modulations $\Omega\le 0.1$  the trajectory
slowly but continually increases its distance to the nearby
part of the adjacent separatrix loop. In the vicinity of the
hyperbolic point the solutions
(\ref{equation:plinear}),(\ref{equation:qlinear}) with $\Omega\le 0.1$
reflect this behavior.

\begin{figure}
\includegraphics[scale=0.5]{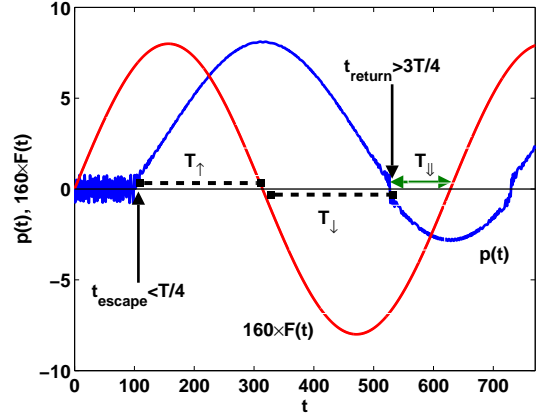}
\caption{(Color online) Illustration of the enhanced motion at work
for directed transport. Shown is the time evolution of the momentum
$p$ and the corresponding force term $F(t)=F\,\sin(\Omega t)$ with
amplitude $F=0.05$. For the sake of comparison $F(t)$ is multiplied by a factor of $160$.
The remaining parameter values are $\Omega=0.01$ and $\Theta_0=0$.
Further details are discussed in the text.}
\label{fig:mechanism.fig}
\end{figure}

\section{Momentum growth}\label{sec:ratchet}

Here we consider the mechanism for the speed-up of a
particle that has escaped from the separatrix and derive an expression for the
effective gain in momentum.

\subsection*{Forced speed-up of particles}
To gain better insight into the mechanism that causes the speed-up
of the particle motion and the transition from librations to
rotations we depict in Fig.~\ref{fig:mechanism.fig} the time
evolution of the momentum for the driven dynamics together with
the corresponding force $F(t)=-\partial U_1/\partial q= F\sin
(\Omega \,t)$ with $\Omega=0.01$. Up to times $t \lesssim 132$ the
trajectory performs librations in the potential well. During the
evolution the particle adiabatically increases its momentum so
that with each turn in the potential it approaches closer  the
saddle point. Eventually, a separatrix crossing takes place when
the trajectory is in the vicinity of the hyperbolic point where
the motion becomes highly irregular, due to the intricate dynamics
connected with the transversal intersections of the invariant
manifolds \cite{Wiggins,Guckenheimer}. Subsequently to such an
escape, the motion proceeds along the direction of the unstable
manifold of the hyperbolic equilibrium as further illustrated in
Fig.~\ref{fig:pqesc.fig}. Most importantly, the separatrix
crossing occurs at an instant of time $t_{escape}<T/4$, for which
the negative inclination has not yet reached its maximum. During
the ongoing particle motion towards the right the momentum
variable performs small-amplitude oscillations, while its average
value grows monotonically. Eventually, the largest value
$p=\sqrt{2/\pi}$ on the unperturbed separatrix (indicated by the
upper straight line in Fig.~\ref{fig:pqesc.fig}) is exceeded. As
the inclination of the potential assumes increasingly negative
values the particles' momentum is raised further. At the end of
the enhancement period, designated by
$T_{\uparrow}=T/2-t_{escape}$ in Fig.~\ref{fig:mechanism.fig}, the
momentum has grown to a maximum value of $p_{max}=8$.
Subsequently, when for $t>T/2$ the inclination of the potential is
positive, the particle is in the phase of depreciation, denoted by
$T_{\downarrow}$. Then, the momentum is reduced steadily and
reverts to zero value at an instant of time
$t_{return}=T/2+T_{\downarrow}>3T/4$. Due to symmetry, one obtains
that $T_{\uparrow}= T_{\downarrow}$.

However, after reversal, i.e. when $p<0$, there remains only
comparatively little time, namely $T_{\Downarrow}=T-t_{return}<T/4$,
during which the motion to the left is enhanced. Consequently, the
asymmetry in the enhancement and depreciation phases, viz. the fact
that  $T/2-t_{escape}
> T/4> T-t_{return}$, serves for a rather long period of
rightwards motion compared to the leftwards  motion. Therefore the
effective motion of the particle proceeds to the right.

\subsection*{Momentum gain}

In fact the momentum gain for a  particle exerted to the force
$F(t)=F\sin(\Omega t+\Theta_0)$ can be estimated as follows: Without
loss of generality we consider phases in the interval $0\le \Theta_0
\le \pi/2$ leading to large motion towards positive momenta if the
particles escape at instants of time $t_{escape}$ such that $\Omega
t_{escape}+\Theta_0 \le \pi/2$ is satisfied. Then there remains the
time interval $ t_{escape}<t\le (\pi-\Theta_0)/\Omega$ during which
the particle still experiences an enhancement in the right
direction. For times $ (\pi-\Theta_0)/\Omega<t\le
(2\pi-\Theta_0)/\Omega$ the force acts in the opposite direction. In
particular for $t_{return}\le t\le (2\pi-\Theta_0)/\Omega$ the
momentum evolves with its sign reversed compared to the previous
enhancement period.  One obtains then
\begin{eqnarray}
\Delta p&=& \left(\int_{t_{escape}}^{(\pi-\Theta_0)/\Omega}+\int_{t_{return}}^{(2\pi-\Theta_0)/\Omega}\right)d\tau\dot{p}\nonumber\\&=&
\left(\int_{t_{escape}}^{(\pi-\Theta_0)/\Omega}+\int_{t_{return}}^{(2\pi-\Theta_0)/\Omega}\right) \nonumber\\
&\times& d\tau\left[-\sin(2\pi
q)+F\sin(\Omega\tau+\Theta_0)\right]\,.
\end{eqnarray}
Due to symmetry, it holds that $t_{return}=(2\pi-\Theta_0)/\Omega-t_{escape}$.
Furthermore, for small $\Omega$ the oscillating part connected with
the first term in the integral averages to zero on the time scale
$t_{escape}\le t \le (2\pi-\Theta_0)/\Omega$ and we find
\begin{equation}
\Delta p= 2\frac{F}{\Omega}\cos(\Omega\, t_{escape}+\Theta_0)\,.\label{equation:pmax}
\end{equation}
In general, the smaller $\Omega$ the higher is the gain in momentum
(see also \cite{Soskin}).
In principle, for a sufficiently small frequency $\Omega$ the gain can become arbitrarily large.
$\Delta p$ is non-negative only if
\begin{equation}
t_{escape} < \frac{\frac{\pi}{2}-\Theta_0}{\Omega}\,.
\end{equation}
Integrating over the interval of
escape times  with the phase $\Theta_0$ being held fixed yields the integrated momentum gain
\begin{eqnarray}
<{\Delta p}>&=&\int_{0}^{(\pi/2-\Theta_0)/\Omega}dt_{escape}\Delta p(t_{escape})\nonumber\\
&=&2\frac{F}{\Omega^2}[1-\sin \Theta_0]\,.
\end{eqnarray}
In conclusion, the integral momentum gain is at its maximum at
$\Theta_0=0$ and diminishes monotonically towards zero at
$\Theta_0 =\pi/2$, underlining the vital (symmetry-breaking) role
of the initial phase $\Theta_0$ for the enhancement process (see
also \cite{Yevtushenko}).

\section{Ensemble Dynamics And Current}\label{sec:current}

In this section we investigate the behavior of the driving induced
current for an ensemble of trapped particles. Their initial
conditions are distributed such that if the potential had the
static inclination $F$ the associated trapped trajectories cannot
cross the corresponding separatrix loop. Likewise, as done in
section \ref{sec:growth} for the chaos-induced detrapping, we
apply a time-periodic modulation of the inclination where the
potential is initially  non-inclined, i.e. we use a fixed initial
phase $\Theta_0=0$. The main objective is to demonstrate  that for
imposed slow modulations a preferred direction of motion emerges
although
 on average the potential landscape remains unbiased.
\begin{figure}
\includegraphics[scale=0.5]{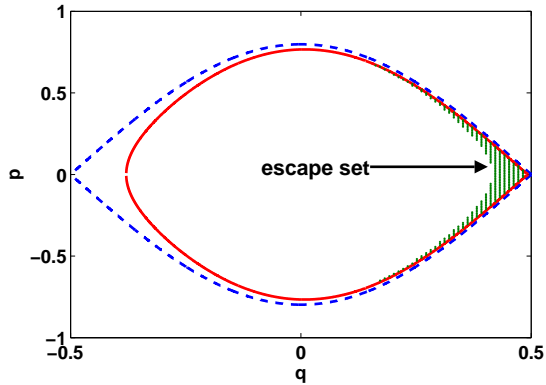}
\caption{(Color online) The separatrix loop of the non-inclined
potential (dashed line) and for the inclined potential due to a static
 force set at $F=0.05$ (solid line). The escape set comprising
those initial conditions that lead to escape out from the solid
separatrix loop is given by the hatched area (in green online).}
\label{fig:escapeset}
\end{figure}

\subsection{Escape times}
First, we identified the escape set, comprising of all those initial
conditions that are contained in the interior of the static
separatrix loop that lead to chaos-induced escape. To this end, the
separatrix loop has been populated densely with points corresponding
to initial conditions for the dynamics of the periodically driven
system. The resulting escape set is displayed in
Fig.~\ref{fig:escapeset} by the hatched area.
\begin{figure}
\includegraphics[scale=0.5]{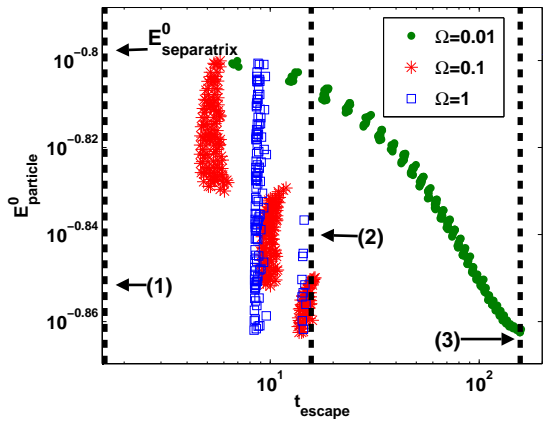}
\caption{ (Color online) The relation between the particle energy $E_{particle}^0$
and the escape time for three different frequencies as indicated in the
plot. The vertical dashed lines marked as $(1)$,$(2)$ and $(3)$ are at the
positions of the enhancement boundary $T/4=\pi/(2\Omega)$ for
$\Omega=1$, marked by $(1)$, $\Omega=0.1$, indicated as $(2)$, and
$\Omega=0.01$, marked as $(3)$, respectively.
The remaining parameter values are $F=0.05$ and $\Theta_0=0$.} \label{fig:esctimer}
\end{figure}
We recall that escape, i.e. the  detrapping, requires a separatrix
crossing. Apparently, the moment of the first separatrix crossing,
i.e.  the trapping-detrapping transition, governs the efficiency of
the enhancement process, as discussed in section
\ref{sec:growth}. Of great importance for the speed-up of an
escaped particle  is that the first separatrix crossing occurs at
instants of time $t_{escape}<T/4=\pi/(2\Omega)$, which determines
the boundary for enhancement. More specifically, according to
Eq.\,(\ref{equation:pmax}), the length of the enhancement period
determines the growth in momentum and thus the value of the
resulting current.

\begin{figure}
\includegraphics[scale=0.35]{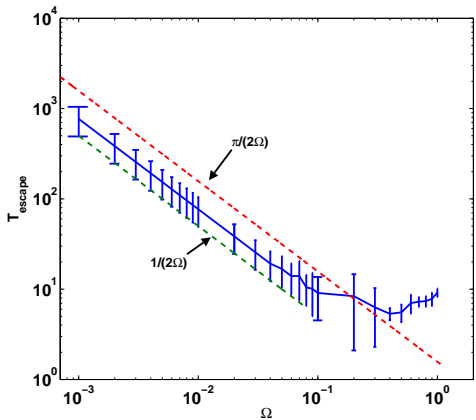}
\caption{(Color online) The mean escape time {\it vs.} the angular
driving frequency is plotted with error bars. The lower and upper
dashed lines correspond to $T_{min}= T/4\pi= 1/(2\Omega)$ and
$T/4=\pi/(2\Omega)$, respectively. The number of particles
emanating from the escape set shown in Fig.~\ref{fig:escapeset} is
taken as $N=512$. The remaining parameter values are $F=0.05$ and
$\Theta_0=0$.} \label{fig:Tesc}
\end{figure}

From Fig.~\ref{fig:esctimer} we  deduce  the corresponding time
scale for escape beyond the separatrix. While for the largest
angular driving frequency $\Omega =1$ the escape takes place at
times far beyond the boundary, marked as $(1)$ in Fig.
~\ref{fig:esctimer}, we note that for an intermediate angular
frequency at $\Omega=0.1$ almost all escape events occur before the
corresponding enhancement boundary $T/4$, marked as $(2)$. For slow
driving at $\Omega =0.01$, however,  the scales of $t_{escape}$
stretch over a wide range, but practically  all escape events do
occur at $t_{escape} < T/4$, marked as $(3)$. Moreover, due to the
irregular nature of the underlying dynamics, for an ensemble of
escaping particles with distributed initial conditions the moments
of their corresponding first separatrix crossing depend sensitively
on the initial conditions. Trajectories starting out very close to
the hyperbolic point are the first to escape, whereas for those
initial conditions which are located  away from the separatrix
considerable time passes until escape takes place. From
Fig.~\ref{fig:Tesc}, depicting the mean escape time versus the
angular driving frequency  $\Omega$, we infer that for low angular
frequencies $\Omega \lesssim 0.04$ the escape times are sufficiently
smaller than $T/4$. In contrast,   for $\Omega \gtrsim 0.04$ at
least the upper error bars of the escape times are of the order of,
or higher, than $T/4$. This  hampers a pronounced enhancement and
in turn tends to suppress the directed flow.
Connecting the lower error bars of the mean escape time by a line we find that the
latter is very well fitted by the expression $T_{min}=1/(2\Omega)$. Since this line
is parallel to the one of the mean escape time the latter obeys the $1/\Omega$-dependence as well.

\subsection{Averaged directed flow}
\begin{figure}
\includegraphics[scale=0.5]{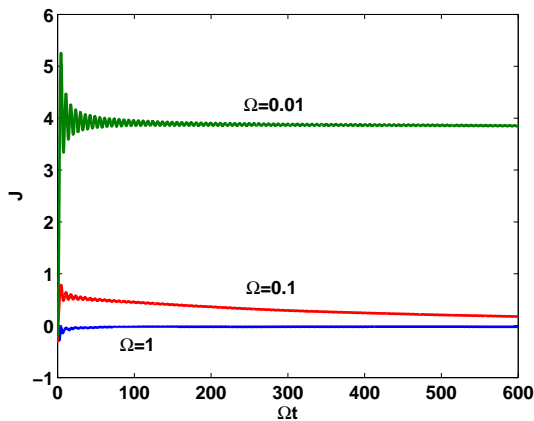}
\caption{(Color online) Temporal behavior of the directed current
defined in Eq.\,(\ref{equation:current}) for three different
angular driving  frequencies of the inclination-modulation force
as labeled correspondingly on the graphs. The amplitude and the
initial phase of the external force are set at  $F=0.05$ and
$\Theta_0=0$, respectively. The number of particles is $N=512$.}
\label{fig:current}
\end{figure}

With regard to directed flow we note that for $\Omega \lesssim
0.04$ we find that for the entire escape set the associated
trajectories move in the {\it same} direction over an extended
period of time. Thus, the generation of a huge, uni-directional
current becomes possible. We define the directed current as the
time average of the ensemble averaged  momentum, i.e.,
\begin{equation}
J=\frac{1}{T_{s}}\,\int_{0}^{T_s}dt<p(t)>\,,\label{equation:current}
\end{equation}
with the ensemble average given by
\begin{equation}
<p(t)>=\frac{1}{N}\sum_{n=1}^{N}(p_n(t)-p_n(0))\label{equation:meanp}\,.
\end{equation}
Here,  $N$ denotes the number of particles constituting the
ensemble. In Fig.~\ref{fig:current} we depict the time evolution of
the directed current $J$ for moderate, intermediate and very slow
modulations.
The time is scaled according to $t\rightarrow \Omega t$; thus
assuring that for a given simulation time $T_s$ the number of
oscillation periods of the external force is the same regardless
of the value of the used angular frequency $\Omega$. At a moderate
driving frequency of $\Omega=1$ the average flow (numerically)
practically vanishes asymptotically on the displayed time scale,  whereas for $\Omega=0.1$
the decay towards  zero  proceeds  slower. This is in compliance with the findings in \cite{Yevtushenko}.

In distinct contrast, however, for extreme slow  driving at $\Omega=0.01$,
the mean momentum assumes a quasi-stationary regime of
considerable large size. We note that our observed averaged
directed current persists over  a long time interval
that agrees well with the explicit estimate for the duration, $t_f$ of
unidirectional motion given in \cite{Soskin}, which
adopted to our system notation reads
\begin{equation}
 t_f\sim\frac{1}{\sqrt{2\pi}}\,\frac{(\sqrt{2\pi}/\Omega)^5}
{\ln^4[\sqrt{2\pi}/(F\,\Omega)]}\,,
\end{equation}
yielding for $\Omega=10^{-2}$ and $F=0.05$ a figure of the order
of $t_f\sim 10^8$. Plotting the directed current as as a function
of the angular driving  frequency $\Omega$ one notices a strong
decay of the current with increasing angular driving frequency
(not shown here). The resulting $1/\Omega$ dependence of the
current corroborates with the expression for the momentum gain
given in Eq.\,(\ref{equation:pmax}); see also in Ref.
\cite{Yevtushenko},\cite{Soskin}, providing intuitive arguments
that support the strong enhancement of the chaotic transport in
space.

\section{Conclusion}

In this work we have investigated the dissipation-less,
time-dependent driven Hamiltonian dynamics of particles evolving
in a symmetric, spatially periodic potential whose inclination is
temporally varied periodically by an external ac-force.

We have focused interest on the generation of a directed flow of
ensembles particles which are trapped in the interior of a
potential well. The choice of the initial starting values is
subjected to the condition that for a static inclination the
trapped particles cannot escape from a potential well. Then, the
only possibility left for escape from the potential well is due to
the chaotic dynamics which arises in the system dynamics due to
the  time-dependent forcing term. In fact, upon applying the
time-periodic modulation of the inclination, trajectories that
become embraced by the developing chaotic layer around the broken
separatrix may cross the latter. However, for fast and
intermediate modulation frequency there results no substantial
directed flow. This is so, because the trajectories frequently
cross and re-cross  the separatrix corresponding to leaving and
re-entering the adjacent wells of the potential wherein the
particles dwell.

For adiabatic inclination modulations  we have demonstrated that
for all initial conditions contained in the escape set, motion
takes place in a unique direction that is controlled by the phase
of the modulation term. It has been shown that the slower the
modulation the larger is the gain in momentum of the escaped
particles and thus the emerging asymptotic current that obeys a
$1/\Omega$ dependence.

Concerning an explanation of this phenomenon it seems that the
cantori, partitioning the nonuniform chaotic layer, are the less
leaky the smaller the modulation frequency $\Omega$. Research
regarding the modulation frequency dependence of the sticking times
to the boundary of regular islands is in progress
\cite{preparation1}. The cantori form almost impenetrable barriers
that confine trajectories for a very long but {\it transient}
period. One should remark that eventually this transient period of
directed motion terminates because the trajectory escapes through
one of the holes in the cantori and accesses other regions of the
chaotic layer. Therefore the motion does not necessarily proceed
unidirectionally: Unless the trajectory gets captured by ballistic
channels it itinerates within the chaotic layer going along with
changes of the direction of motion.

Suitable physical systems that come to mind to  experimentally test
our findings are periodically driven cold atom optical lattices, as
recently fabricated  in studying dissipative, classical ratchet
dynamics \cite{Renzoni1,Renzoni2}. In order to verify our
multi-facetted findings  of  directed, dissipation-less transport
the use of off-resonant, far detuned laser beams is required;
thus minimizing the dissipation in these cold atom set-ups: A
scenario proposed in different context also for Hamiltonian quantum
ratchets in Refs. \cite{Gong,Denisov07}.
Moreover, the particle trapping-detrapping transitions induced by time-dependent modulations
of the potential
as described in this manuscript can also be applied in fluid dynamics to design particle traps
in incompressible open flows as discussed in \cite{Benczik}.

As an interesting extension of the present work we currently  engage
in studying this dissipation-less enhancement effect for  coupled
nonlinear systems which are composed of a chain of interacting
particles \cite{preparation}.

\vspace{0.5cm} \centerline{\large{\bf Acknowledgments}}

\noindent This research was supported by SFB 555 and the VW
Foundation Projects I/80425 (L.Sch.-G.) and I/80424 (P.H.). We also
acknowledge very insightful  and constructive discussions with S.
Denisov.

\end{document}